\documentclass[pre,groupaddress,twocolumn,showpacs]{revtex4}
\usepackage{epsfig,amsmath,amssymb,graphics,color,calc}

\newcommand{\be}{\begin{equation}}
\newcommand{\ee}{\end{equation}}
\newcommand{\ba}{\begin{eqnarray}}
\newcommand{\ea}{\end{eqnarray}}

\begin{document}

\title{Revisiting the slow dynamics of 
a silica melt using Monte Carlo simulations} 

\author{Ludovic Berthier}
\affiliation{Joint Theory Institute, Argonne National Laboratory and
University of Chicago,
5640 S. Ellis Av., Chicago, Il 60637}

\altaffiliation{Permanent address: Laboratoire des Collo{\"\i}des, Verres
et Nanomat{\'e}riaux, UMR 5587, Universit{\'e} Montpellier II and CNRS,
34095 Montpellier, France}

\date{\today}

\begin{abstract}
We implement a standard Monte Carlo algorithm to 
study the slow, equilibrium dynamics of a silica melt in a wide 
temperature regime, from 6100~K down to 2750~K. 
We find that the average dynamical
behaviour of the system is in quantitative agreement 
with results obtained from molecular dynamics simulations,
at least in the long-time regime corresponding to the 
alpha-relaxation. By contrast, the strong thermal vibrations
related to the Boson peak present at short times in molecular dynamics  
are efficiently suppressed by the Monte Carlo algorithm.
This allows us to reconsider silica dynamics
in the context of mode-coupling theory, because several shortcomings of the 
theory were previously attributed to thermal vibrations. 
A mode-coupling theory analysis of our data 
is qualitatively correct, but quantitative tests of the theory fail, 
raising doubts about the very existence of an avoided singularity
in this system. 
We discuss the emergence of dynamic heterogeneity and report detailed
measurements of a decoupling between translational diffusion and structural 
relaxation, and of a growing four-point dynamic susceptibility.
Dynamic heterogeneity appears to be less pronounced 
than in more fragile glass-forming models, but not of a qualitatively 
different nature.
\end{abstract}

\pacs{02.70.Ns, 64.70.Pf, 05.20.Jj}

%61.20.Lc  Time-dependent properties; relaxation (for glass transitions, see 64.70.Pf)

%02.70.Ns  Molecular dynamics and particle methods

%05.20.Jj  Statistical mechanics of classical fluids (see also 47.10.-g General theory in fluid dynamics)

\maketitle

\section{Introduction}

Numerical simulations play a major role among 
studies of the glass transition
since, unlike in experimental works,
the individual motion of a large number of particles 
can be followed at all times~\cite{hans}. 
Computer simulations usually study Newtonian dynamics (ND) by solving 
a discretized version of Newton's equations
for a given interaction  between particles~\cite{at}. 
Although this is the most appropriate 
dynamics to study molecular liquids, it can be interesting 
to consider alternative dynamics that are not deterministic, 
or which do not conserve the energy. 
In colloidal glasses and physical gels, for instance, 
the particles undergo Brownian motion arising
from collisions with molecules in the solvent, and a stochastic 
dynamics is more appropriate~\cite{at}. Theoretical considerations
might also suggest the study of different sorts of dynamics
for a given interaction between particles, for instance to assess
the role of conservation laws~\cite{previous,I,II} 
and structural information~\cite{mct,mctcolloid}.
Of course, if a given dynamics satisfies detailed balance with respect to 
the Boltzmann distribution, all structural quantities remain unchanged, 
but the resulting dynamical behaviour might be very different.
In this paper, we study the relaxation dynamics 
of a commonly used theoretical model for  silica
using Monte Carlo simulations and compare the results with previous ND 
studies for both the averaged dynamical behaviour and
the spatially heterogeneous dynamics of this system.

Several papers have studied in detail the influence of the chosen
microscopic dynamics on the dynamical behaviour in a simple
glass-former, namely a binary mixture of Lennard-Jones 
particles~\cite{hans,KA}.
Gleim, Kob and Binder~\cite{gleim} studied Stochastic Dynamics 
where a friction term and a random noise 
are added to Newton's equations, the amplitude of both terms 
being related by a fluctuation-dissipation theorem.
Szamel and Flenner~\cite{szamel2} used Brownian
dynamics, in which there are no momenta, 
and positions evolve with a Langevin dynamics. 
Berthier and Kob~\cite{ljmc} employed Monte Carlo dynamics, 
where the potential energy between two configurations 
is used to accept or reject a trial move.
The equivalence between these three stochastic dynamics
and the originally studied ND was established at the level of the averaged 
dynamical behaviour, except at very short times where differences 
are indeed expected. However, important
differences were found when dynamic fluctuations 
were considered, even in the 
long-time regime comprising the alpha-relaxation~\cite{I,II,ljmc}.

Silica, the material studied in the present work, 
is different from the previously considered Lennard-Jones 
case in many aspects which all motivate our Monte Carlo study of
the Beest, Kramer and van Santen (BKS) model for 
silica~\cite{beest90}.
First, the BKS model was devised to represent a real material, making
our conclusions more directly applicable to experiments. Second, 
the temperature evolution of relaxation times 
is well-described by a simple Arrhenius law at low temperatures, typical 
of strong glass-formers, which are commonly believed 
to belong to a somewhat different class of materials, so that
qualitative differences might be expected with more fragile, 
super-Arrheniusly relaxing materials.
Third, the onset of slow dynamics in fragile materials 
is often said to be accurately described by  the application
of mode-coupling theory, at least over 
an intermediate window of 2 to 3 decades of relaxation
times~\cite{mct}. Mode-coupling theory (MCT) formulates in particular a series 
of quantitative predictions regarding the time, spatial, and temperature
dependences of dynamic correlators. In the case of
silica melts, previous analysis reported evidence in favour of a narrower 
temperature regime where MCT can be applied, but the test
of several theoretical predictions was either 
seriously affected, or even made impossible
by the presence of strong short-time thermal vibrations 
related to the Boson peak in this material~\cite{hk}. 
These vibrations affect
the time dependence of the correlators much more strongly 
in silica than in Lennard-Jones systems, which constitutes
a fourth difference between the two systems. Using
Monte Carlo simulations we shall therefore be able to 
revisit the MCT analysis performed in Refs.~\cite{hk}. 
Fifth, while detailed analysis of dynamic heterogeneity
is available for fragile materials, a comparatively smaller 
amount of data is available for strong materials~\cite{I,II,vogel,teboul}, 
and we shall 
therefore investigate issues that have not been addressed in previous
work.

The paper is organized as follows. In Sec.~\ref{mc} we give 
details about the simulation technique and compare its efficiency 
to previously studied dynamics. In Sec.~\ref{results} we present 
our numerical results about the averaged dynamics of silica in 
Monte Carlo simulations, while Sec.~\ref{dh} 
deals with aspects related to dynamic heterogeneity.
Finally, Sec.~\ref{conclusion} concludes the paper.  

\section{Simulating silica using Monte Carlo dynamics}
\label{mc}

Our aim is to study a non-Newtonian dynamics of the glass-former 
silica, SiO$_2$. 
Therefore, we must first choose a reliable model to describe the 
interactions in this two-component system made of 
Si and O atoms, and then design a specific stochastic
dynamics which we require to be efficient and to
yield the same static properties as Newtonian dynamics.
 
Various simulations have shown that a reliable pair potential  
to study silica in computer simulations is the one 
proposed by BKS~\cite{beest90,hk,vogel,teboul,bks_sim,bks_sim2,voll}. 
The functional form of the BKS potential is 
\begin{equation} 
\phi_{\alpha \beta}^{\rm BKS}(r)= 
\frac{q_{\alpha} q_{\beta} e^2}{r} + 
A_{\alpha \beta} \exp\left(-B_{\alpha \beta}r\right) - 
\frac{C_{\alpha \beta}}{r^6}, 
\label{pp} 
\end{equation} 
where $\alpha, \beta \in [{\rm Si}, {\rm O}]$ and $r$ is the distance 
between the atoms of type $\alpha$ and $\beta$.  The values of the 
constants $q_{\alpha}, q_{\beta}, A_{\alpha \beta}, B_{\alpha \beta}$, 
and $C_{\alpha \beta}$ can be found in Ref.~\cite{beest90}. For the 
sake of computational efficiency the short range part of the potential 
was truncated and shifted at 5.5~\AA. This truncation also has the 
benefit of improving the agreement between simulation and experiment 
with regard to the density of the amorphous glass at low temperatures. 
The system investigated has $N_{\rm Si} = 336$ and $N_{\rm O}=672$ 
atoms in a cubic box with fixed size $L=24.23$~\AA, so 
that the density is $\rho = 2.37$ g/cm$^3$, close to the 
experimentally measured density at atmospheric pressure
of 2.2~g/cm$^3$~\cite{density}.

Once the pair interaction is chosen, we have to decide what stochastic dynamics
to implement. Previous studies in a Lennard-Jones system concluded 
that among Monte Carlo (MC), Stochastic Dynamics (SD)
and Brownian Dynamics (BD), 
MC was by far the most efficient algorithm because
relatively larger incremental steps can be used while 
maintaining detailed balance, which is impossible for 
SD and BD where very small discretized timesteps are needed 
to maintain the fluctuation-dissipation relation between 
noise and friction terms~\cite{ljmc}. Given the generality of this argument, 
it should carry over to silica, and we decided to implement MC
dynamics to the BKS model. An additional justification for our choice stems
from unpublished results by Horbach and Kob who performed preliminary 
investigations of the SD of BKS silica~\cite{phdjurgen}. 
Using a friction term similar
in magnitude to the one used in Lennard-Jones simulations was however not
enough to efficiently suppress short-time elastic vibrations. 
Using an even larger friction term would probably
damp these vibrations, but would also make the simulation 
impractically slow. 

A standard Monte Carlo dynamics~\cite{at} 
for the pair potential in Eq.~(\ref{pp}) should proceed as follows.  
In an elementary MC move,
a particle, $i$, located at the position ${\bf r}_i$  
is chosen at random. The energy cost, $\Delta E_i$, to move 
particle $i$ from position ${\bf r}_i$ to a new, 
trial position ${\bf r}_i + \delta {\bf r}$ is evaluated,   
$\delta {\bf r}$ being a random vector comprised in a cube 
of linear length $\delta_{\rm max}$ centered around the origin.
The Metropolis acceptance rate, $p = {\rm min} 
(1, e^{-\beta \Delta E_i /k_B})$, where 
$\beta =1/T$ is the inverse temperature,
is then  used to decide whether the move is accepted or rejected. 
In the following, one Monte Carlo timestep represents 
$N=N_{\rm Si} + N_{\rm O}$ 
attempts to make such an elementary move, and timescales
are reported in this unit. Temperature will be expressed in Kelvin.
Monte Carlo simulations can of course be made even more efficient by 
implementing for instance swaps between particles, or using 
parallel tempering. The dynamical behaviour, however, is then strongly
affected by such non-physical moves
and only equilibrium thermodynamics can be studied. Since 
we want to conserve a physically realistic 
dynamics, we cannot use such improved schemes.

An additional difficulty with Eq.~(\ref{pp}) as compared to 
Lennard-Jones systems is the Coulombic interaction in the first term.
Such a long-range interaction means that the evaluation of the energy 
difference $\Delta E_i$ needed to move particle $i$ in an elementary 
MC step requires a sum over every particles $j \neq i$, and over 
their repeated images due to periodic boundary conditions. 
Of course the sum
can be efficiently evaluated using Ewald summations techniques,
as is commonly employed in ND simulations~\cite{at}. We note, however, 
that Ewald techniques are better suited for ND than for MC since
in ND the positions and velocities of all particles are simultaneously
updated so that the Ewald summation is performed  once to update 
all particles. 
In MC simulations,
each single move requires its own Ewald summation, and this remains
computationally very costly. 

For the BKS potential
in Eq.~(\ref{pp}) it was recently shown that a simple truncation 
can be performed which makes the range of the Coulombic interaction term
finite~\cite{carre}. 
Detailed ND simulations have shown that in the range of temperatures
presently accessible to computer experiments,
no difference can be detected between 
the finite range and the infinite range versions of the BKS potential
for a wide variety of static and dynamic properties.
Therefore we build on this work and make the replacement~\cite{carre}
\be
\frac{1}{r} \to \left( \frac{1}{r} - \frac{1}{r_c} \right)
+ \frac{1}{r_c^2} (r-r_c), \quad {\rm for} \,\, r \leq r_c,
\ee 
while $1/r \to 0$ for $r>r_c$. This amounts to smoothly truncating
the potential at a finite range, $r_c$, maintaining both
energy and forces continuous at the cutoff $r=r_c$.  
The physical motivation for this form of the truncation was given
by Wolf~\cite{wolf}, and discussed in several 
more recent papers~\cite{carre,wolf2}.
Following Ref.~\cite{carre}, we fix $r_c=10.14$~\AA. 
Once the potential is truncated, MC simulations become 
much more efficient, and much simpler to implement. Furthermore, 
this will allow us to perform
detailed comparisons of the dynamics of BKS model of silica 
where the Wolf truncation is used for both ND and MC in the 
very same manner, so that any difference between the two sets of data can be 
safely assigned to the change of microscopic dynamics alone, while reference
to earlier work done using Ewald summations is still quantitatively 
meaningful.

\begin{figure}
\begin{center}
\psfig{file=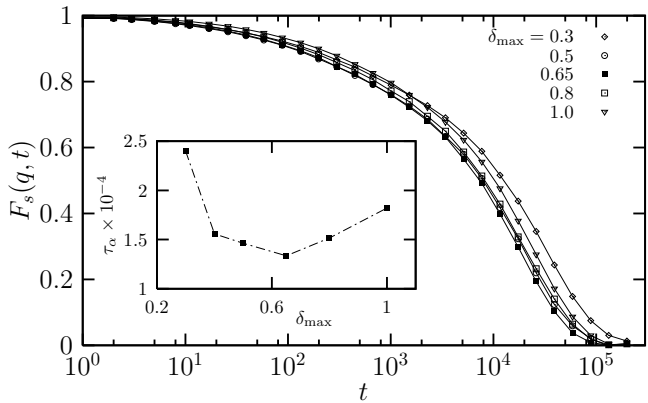,width=8.5cm}
\end{center}
\caption{\label{delta} Self-intermediate scattering function
for silicon, Eq.~(\ref{self}), at $T=4000$~K and $|{\bf q}|=1.7$~\AA$^{-1}$ 
for various values of $\delta_{\rm max}$.
Inset: The evolution of the relaxation time with 
$\delta_{\rm max}$ unambiguously defines an optimal value
$\delta_{\rm max} \approx 0.65$~\AA~ 
for efficient Monte Carlo simulations.}
\end{figure}

The one degree of freedom that remains to be fixed is $\delta_{\rm max}$,
which determines the average lengthscale of elementary moves. 
If chosen too small, energy costs are very small and most of the moves
are accepted, but the dynamics is very slow because it takes a long 
time for particles to diffuse over the long distances needed to relax
the system.
On the other hand too large displacements will on average be very costly 
in energy
and acceptance rates can become prohibitively small. We seek a compromise 
between these two extremes by monitoring the 
dynamics at a moderately low temperature, $T=4000$~K, for several  
values of $\delta_{\rm max}$. 
As a most sensitive indicator of the relaxational behaviour, we measure the 
contribution from the specie $\alpha$ ($\alpha=$ Si, O) to the 
self-intermediate scattering function defined by
\begin{equation} F_s(q,t) = 
\left\langle \frac{1}{N_\alpha} \sum_{j=1}^{N_\alpha} e^{i {\bf q} 
\cdot [{\bf r}_j(t) - {\bf r}_j(0)]} \right\rangle. 
\label{self} 
\end{equation}
We make use of rotational invariance to
spherically average over wave vectors of comparable magnitude, 
and present results for $|{\bf q}|=1.7$~\AA$^{-1}$, 
which is the location of the
pre-peak observed in the static structure factor 
$S(q)$ of the liquid. This corresponds to the typical (inverse) size of the
SiO$_4$ tetrahedra. 
In Fig.~\ref{delta} we present our results for $\delta_{\rm max}$ 
values between 0.3 and 1.0~\AA. 
As expected we find that relaxation is slow both at small and large 
values of $\delta_{\rm max}$, and most efficient for intermediate values.  
Interestingly we also note that the overall 
shape of the self-intermediate
scattering function does not sensitively depend 
on $\delta_{\max}$ over this wide range. We can therefore safely fix 
the value of $\delta_{\max}$ based on an efficiency criterium alone.

We define a typical relaxation time, $\tau_\alpha$,  as 
\be
F_s(q,\tau_\alpha) = \frac{1}{e},
\label{deftaualpha}
\ee 
and show its $\delta_{\rm max}$ dependence
in the inset of Fig.~\ref{delta}. A clear minimum is observed
at the optimal value of $\delta_{\max} \approx 0.65$~\AA, which 
we therefore use throughout the rest of this paper.
This distance corresponds 
to a squared displacement of 0.4225~\AA$^2$, which is very close
to the plateau observed at intermediate times in the 
mean-squared displacements (see Fig.~\ref{fs} below). 
This plateau can be taken 
as a rough measurement of the ``cage'' size for the particles, so that
MC simulations are most efficient when the cage is most quickly explored.
This argument and the data in Fig.~\ref{fs} 
suggest that the location of the minimum should only be a weak function of 
temperature, but we have not verified this point in detail. Therefore
we keep the value of $\delta_{\max}$ constant at all temperatures. 
An alternative would be to optimize it at each $T$ and 
then carefully rescale timescales between runs at different temperatures. 

What about the relative efficiency between MC and ND? If we compare 
the relaxation measured at $T=4000$~K, we find
$\tau_\alpha \sim 13400$ Monte Carlo steps, while 
$\tau_\alpha \sim 4.7$~ps for ND. When using a discretized timestep of 
1.6~fs, this means that, when counting in number of integration
timesteps, MC dynamics is $\approx 5$ times slower than 
ND. This result contrasts with the results obtained in 
a Lennard-Jones mixture where MC dynamics was about 
2 times faster than ND~\cite{ljmc}. We attribute this relative loss 
of efficiency to the existence of strong bonds between Si and O atoms
in silica, which have no counterparts in Lennard-Jones systems.
It is obvious that strong bonds are very hard to relax when using
sequential Monte Carlo moves, as recently 
discussed in Ref.~\cite{steve}. 

We have performed simulations at temperatures between 
$T=6100$~K and $T=2750$~K, the latter being smaller than
the fitted mode-coupling temperature, $T_c=3330$~K~\cite{hk}.
For each temperature we have simulated 3 independent samples 
to improve the statistics. Initial configurations 
were taken as the final configurations obtained from 
previous work performed with ND~\cite{carre}, so that production 
runs could be started immediately. For each sample, production 
runs lasted at least $15\tau_\alpha$ (at $T=2750$~K),
much longer for higher temperatures, so that statistical errors 
in our measurements are fairly small. We have performed a few runs for a larger
number of particles, namely $N=8016$ particles, to investigate finite
size effects which are known to be relevant in silica~\cite{hka,fss,fss2},
and the results 
will be discussed in Sec.~\ref{results}.

\section{Analysis of averaged two-time correlators}
\label{results}

In this section we report our results about the time behaviour
of averaged two-time correlators, we compare the Monte Carlo
results to Newtonian dynamics, and we perform a quantitative 
mode-coupling analysis of the data.

\subsection{Intermediate scattering function and mean-squared displacements}

\begin{figure}
\psfig{file=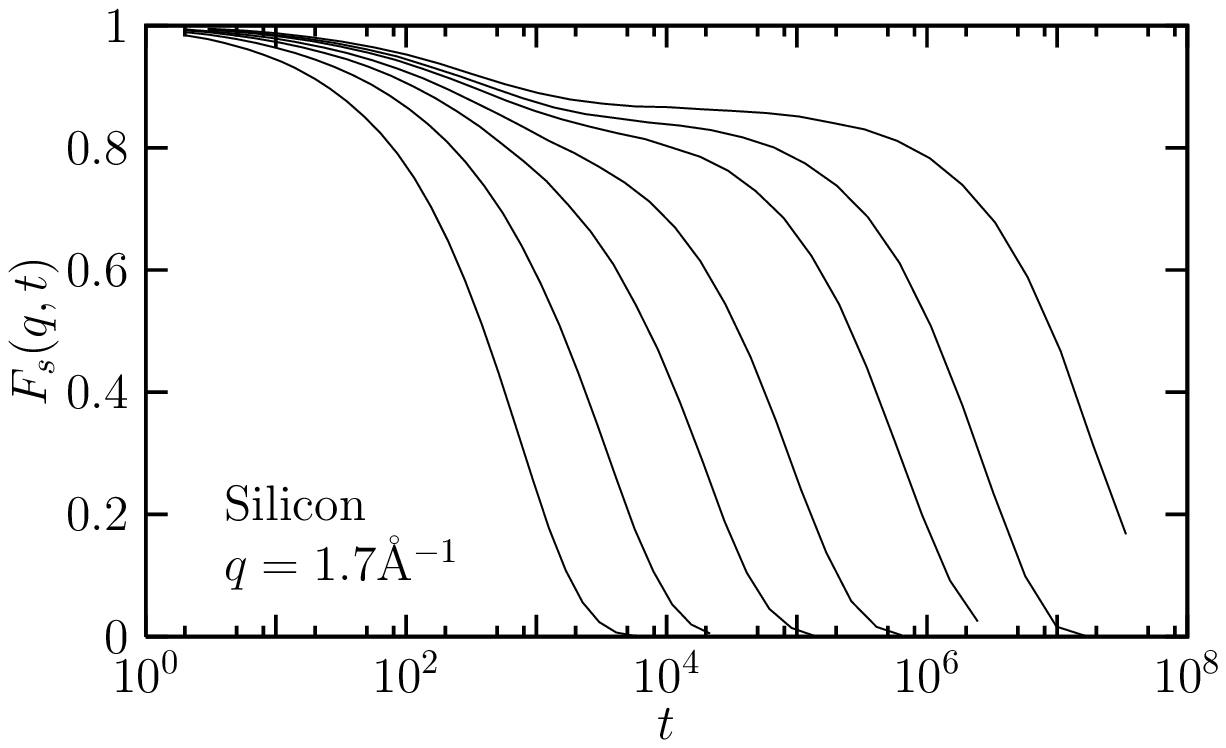,width=8.5cm}
\psfig{file=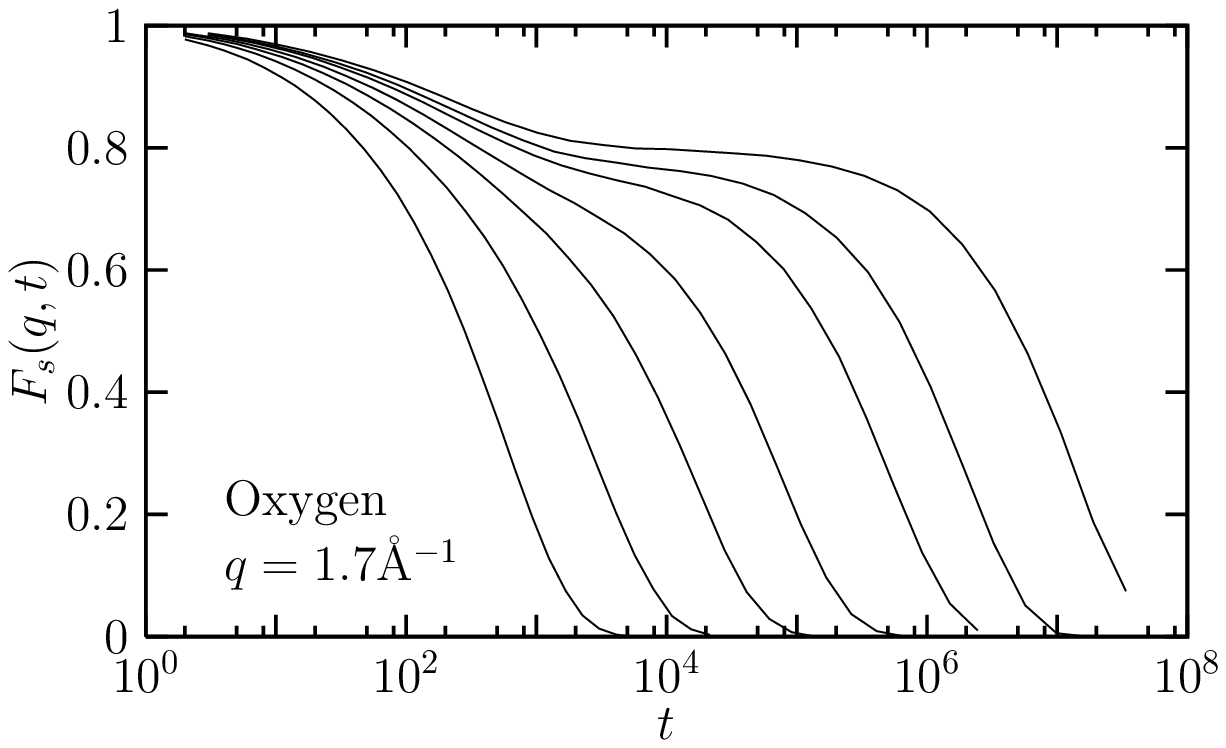,width=8.5cm}
\psfig{file=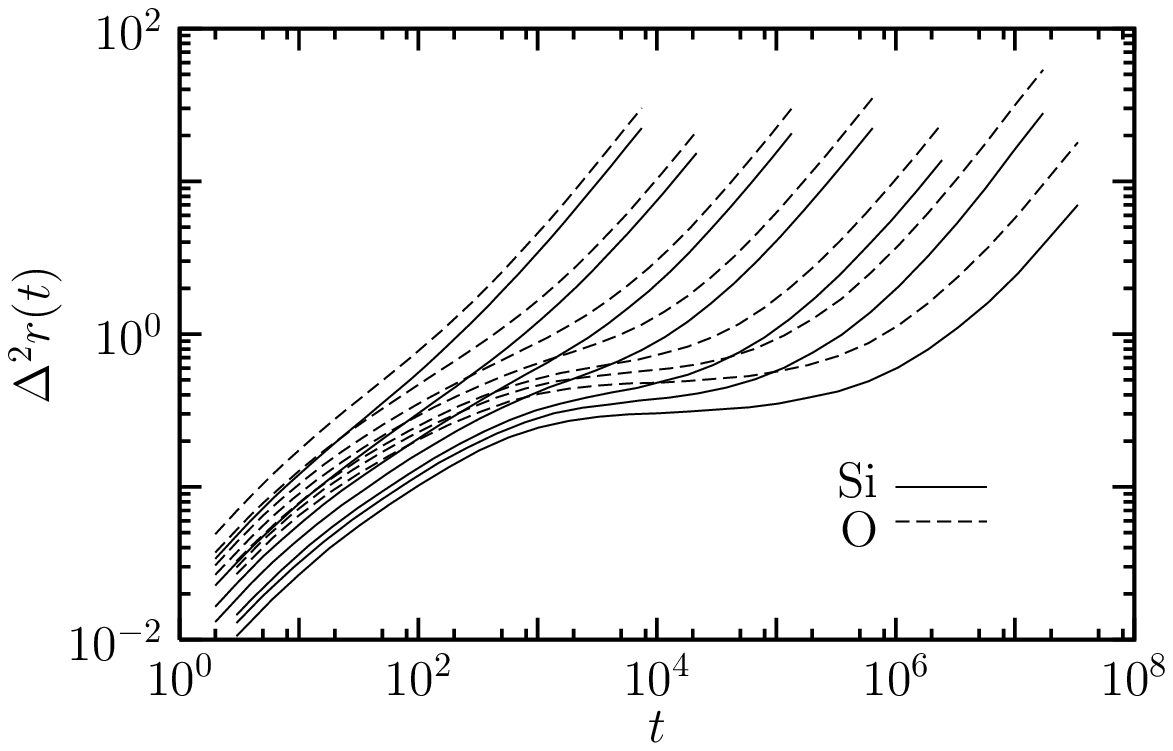,width=8.5cm}
\caption{\label{fs} Top: Self-intermediate scattering function, 
Eq.~(\ref{fs}), for $|{\bf q}|=1.7$~\AA$^{-1}$ and  
temperatures $T=6100$, 4700, 4000, 3580, 3200, 3000, and 2750~K 
(from left to right).
Bottom: Mean-squared displacement, Eq.~(\ref{msd}), for the same 
temperatures in the same order.}
\end{figure}

The self-intermediate scattering function, Eq.~(\ref{self}), 
is shown in Fig.~\ref{fs}
for temperatures decreasing from $T=6100$~K down to $T=2750$~K
for Si and O atoms at $|{\bf q}| = 1.7$~\AA$^{-1}$. 
These curves present well-known features. Dynamics at high temperature 
is fast and has an exponential nature. When temperature is decreased
below $T \approx 4500$~K, a two-step decay, the slower being strongly 
non-exponential, becomes apparent. Upon decreasing 
the temperature further,
the slow process dramatically slows down by about 4 decades, 
while clearly conserving an almost temperature-independent, 
non-exponential shape, as already reported for ND~\cite{hk}.

We also find that the first process, the decay towards
a plateau, slows down considerably when decreasing temperature, although
less dramatically than the slower process. 
The fastest process, called `critical decay' in the language 
of mode-coupling theory~\cite{mct}, is not observed when using 
ND, because it is obscured by strong thermal vibrations
occurring at high frequencies (in the THz range). Clearly, no such vibrations 
are detected in the present results which demonstrates 
our first result: MC simulations very efficiently suppress 
the high-frequency oscillations observed with ND.

Although the plateau seen in $F_s(q,t)$ is commonly
interpreted as `vibrations of a particle within a cage',
the data in Fig.~\ref{fs} discard this view. From direct
visualization of the particles' individual dynamics
it is obvious that vibrations take place in just a few MC
timesteps, while the decay towards the plateau can be 
as long as $10^4$ time units at the lowest temperatures studied here. 
This decay is therefore necessarily more complex, 
most probably cooperative in nature. This interpretation
is supported by recent theoretical studies where a plateau 
is observed in two-time correlators of lattice models where local
vibrations are indeed completely absent~\cite{bethe}. 
A detailed atomistic description 
of this process has not yet been reported, but would 
indeed be very interesting.

Next, we study the mean-squared displacement
defined as 
\begin{equation}
\label{msd}
\Delta^2 r(t) = \frac{1}{N_\alpha} \sum_{i=1}^{N_\alpha} \left\langle 
 |{\bf r}_i(t) - {\bf r}_i(0) |^2 \right\rangle,
\end{equation}
and we present its temperature evolution in Fig.~\ref{fs}, 
for both Si and O atoms. The evolution of $\Delta^2 r(t)$
mirrors that of the self-intermediate scattering function, 
and the development of a two-step relaxation process
is clear from these figures.  
Because we are studying a stochastic dynamics, 
displacements are diffusive at both short and long timescales.
This constitutes an obvious, expected difference between
ND and MC simulations: data clearly cannot match at very small 
times. The goal of the present study is therefore to determine 
whether the dynamics quantitatively match at times where the 
relaxation is not obviously ruled by short-time ballistic/diffusive 
displacements.
 
The plateau observed in $F_s(q,t)$ now translates into a strongly 
sub-diffusive
regime in the mean-squared displacements separating the two diffusive
regimes. At the lowest temperature studied, when $t$ changes by three decades
from $2\times10^2$ to $2\times 10^5$, the mean-squared displacement
of Si
changes by a mere factor 4.6 from 0.16 to 0.074.
Particles are therefore nearly arrested for several decades of times,
before eventually entering the diffusing regime where
the relaxation of the structure of the liquid takes place. 

\subsection{Comparison to Newtonian dynamics}

The previous subsection has shown that the Monte Carlo dynamics
of silica is qualitatively similar to the one reported for ND,
apart at relatively short times where the effect of thermal vibrations 
is efficiently suppressed and the dynamics is diffusive instead
of ballistic. We now compare our results more
quantitatively with the dynamical behaviour observed using ND. 
 
\begin{figure}
\hspace*{-0.2cm}
\psfig{file=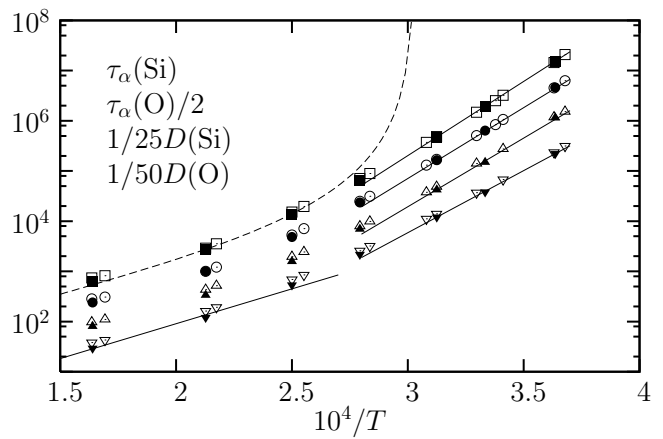,width=8.5cm}
\caption{\label{comp} 
Temperature evolution of the alpha-relaxation time
$\tau_\alpha(T)$ for silicon (squares) and oxygen (circles), 
and inverse self-diffusion constant for silicon (up triangles)
and oxygen (down triangles), vertically shifted 
for clarity. Open symbols are for ND (times rescaled by 
$t_0=0.31$ fs) closed symbols for MC. 
Full lines are Arrhenius fits below $T\approx 3700$~K with
activation 5.86, 5.60, 5.43, and 4.91~eV (from top to bottom).
An Arrhenius fit for high  temperatures is also presented 
for $D$(O) with activation energy 2.76~eV. 
The dashed line is a power law fit, $\tau_\alpha \sim 
(T-T_c)^{-\gamma}$, with $T_c=3330$~K and $\gamma=2.35$.}
\end{figure}

To this end, we compare first the temperature evolution of the 
relaxation times, $\tau_\alpha(T)$, defined in Eq.~(\ref{deftaualpha}),
in Fig.~\ref{comp}. Here, we use a standard 
representation where an Arrhenius slowing down over a 
constant energy barrier $E$, with an attempt frequency $1/\tau_0$,
\be 
\tau_\alpha = \tau_0 \exp \left( \frac{E}{k_B T} \right),
\label{arrhenius}
\ee
appears as a straight line. To compare both sets of data we rescale 
the ND data by a common factor, $t_0 = 0.31$~fs, which 
takes into account the discretization timestep and the 
efficiency difference discussed in the previous section; $t_0$ will
be kept constant throughout this paper.
We find that the temperature 
evolution of the alpha-relaxation time measured in MC simulations
is in complete quantitative agreement with the 
one obtained from ND, over the complete temperature range.
This proves that Monte Carlo techniques can be applied
not only to study static properties of silica, but also its
long-time dynamic properties.

In Fig.~\ref{comp} we also show the temperature evolution
of the self-diffusion constant, defined from the long-time limit
of the mean-squared displacement as 
\begin{equation}
D = \lim_{t \to \infty} \frac{\Delta^2 r(t)}{6 t}.
\end{equation}
The behaviour of the (inverse) 
diffusion constant is qualitatively very close 
to the one of the alpha-relaxation time, and we again find 
that ND and MC dynamics yield results in full quantitative agreement.

As expected for silica, we find that at low temperatures below
$T \approx 3700$~K, relaxation timescales and diffusion constant 
change in an Arrhenius fashion described by Eq.~(\ref{arrhenius}).
We find, however, that the observed activation energies 
display small variations between different observables, from 
5.86~eV for $\tau_\alpha$(Si) to 4.91~eV for $1/D$(0).
These values compare well with previous analysis~\cite{hk}, 
and with experimental
findings~\cite{silicaexp}.

From Fig.~\ref{comp}, it is clear that Arrhenius behaviour is 
obeyed below $T \approx 3700$~K only, while the data bend up
in this representation for higher temperatures. This behaviour 
was interpreted in terms of a fragile to strong behaviour of the 
relaxation timescales in several papers~\cite{hk,ftos1,ftos2}, 
despite the fact that
fragility is usually defined experimentally by considering data
on a much wider temperature window close to the experimental
glass transition.  
To rationalize these findings, Horbach
and Kob analyzed the data using mode-coupling theory 
predictions~\cite{hk}. 
In particular they suggest to fit the temperature 
dependence of $\tau_\alpha$ as
\be 
\tau_\alpha \sim (T-T_c)^{-\gamma},
\label{taumct}
\ee
with $T_c \approx 3330$~K and $\gamma \approx 2.35$. 
This power law fit is also 
presented in Fig.~\ref{comp} as a dashed line. Its domain 
of validity is of about 1 decade, which is significantly less 
than for more fragile materials with super-Arrhenius behaviour
of relaxation timescales~\cite{KA}. 

It is interesting to note that a simpler 
interpretation of this phenomenon could be that this behaviour is nothing
but a smooth crossover from a non-glassy, homogeneous, high-temperature 
behaviour to a glassy, heterogeneous, low temperature 
behaviour, as found in simple models of strong glass-forming 
liquids~\cite{bg}. 
In Fig.~\ref{comp}, we implement this simpler scenario by fitting high
temperature data with an Arrhenius law, as is sometimes  
done in the analysis of experimental data~\cite{gilles}. 
Such a fit works nicely for 
high temperatures, from $T=6100$ to 4700~K, but breaks down 
below $T \approx 4000$~K. 
A physical interpretation for this high-temperature Arrhenius behaviour
was offered in Ref.~\cite{heuer}.
This shows that analyzing silica dynamics
in terms of a simple crossover occurring around 4000~K between two simple
Arrhenius law is indeed a fair description of the data which does not require
invoking a more complex fragile to strong crossover being rationalized by the 
existence of an avoided mode-coupling singularity.  

\begin{figure}
\psfig{file=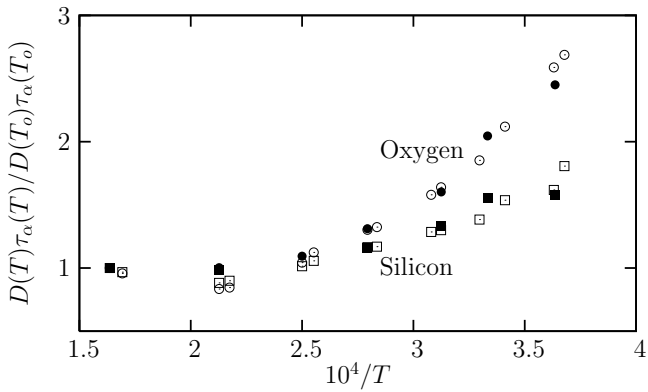,width=8.5cm}
\caption{\label{comp2} 
Decoupling data for oxygen and silicon. We plot the 
product $D \tau_\alpha$ taken from the data shown 
in Fig.~\ref{comp} and normalize the product
by its value at $T_o=4700$~K such that deviations
from 1 indicates non-zero decoupling. 
Open symbols are for ND, closed symbols for MC. Decoupling
is similar for both types of dynamics.}
\end{figure}

The difference found above for the activation energies describing
$\tau_\alpha$ and $1/D$ for both species 
implies that these quantities, although both
devised to capture the temperature evolution of 
single particle displacements have slightly different
temperature  evolutions and are not proportional to one another.
This well-known feature implies the existence of 
a ``decoupling'' between translational 
diffusion and structural relaxation in silica, 
as documented in previous papers~\cite{hk}. 
In Fig.~\ref{comp2} we report the temperature evolution
of the product $D(T) \tau_\alpha(q,T)$ which is a pure constant
for a simply diffusive particle where $\tau_\alpha(q,T) = 1/(q^2 D)$.
We normalize this quantity by its value at $T_o=4700$~K, so 
that any deviations from 1 indicates a non-zero 
decoupling~\cite{berthier,epl}.
As expected we find that the product is not a constant, 
but grows when temperature decreases. Remarkably, although this 
quantity is a much more sensitive probe of the dynamics of the liquid,
its temperature evolution remains quantitatively similar for both ND and 
MC dynamics. This shows that equivalence of the dynamics between
the two algorithms holds at the level of the complete distribution
of particle displacements, even for those tails that
are believed to dictate the observed decoupling. 

In Sec.~\ref{dh}, we shall explore in more detail the 
heterogeneous character of the dynamics of silica,
closely related to the decoupling discussed here. It is however 
interesting to try and infer the amount of decoupling 
predicted for silica at temperatures close to the experimental 
glass transition, $T_g \approx 1450$~K. The glass transition temperature
of BKS silica deduced from extrapolation of viscosity measurements
is close to the experimental one, 
$T_g^{\rm BKS} \approx 1350$~K~\cite{hk}. Extrapolating the 
data in Fig.~\ref{comp2} down to 1400~K predicts 
a decoupling of about 40 for Si dynamics, about 7 for O dynamics. 
The difference between Si and O dynamics was recently 
explained in Ref.~\cite{heuer}, where
it was noted that oxygen diffusion is in fact possible with no 
rearrangement of the tetrahedral structure of silica involved. 
Moreover, it is interesting to note that the amount of decoupling 
found here is smaller than experimental findings in fragile materials
close to their glass transition~\cite{decouplingexp}, 
but is nonetheless clearly 
different from zero. This suggests that 
even strong materials display 
dynamically  heterogeneous dynamics, but its effect seems 
less pronounced than in more fragile materials.

Theoretically, an identical temperature evolution of the
alpha relaxation timescale for MC and ND 
is an important prediction of mode-coupling theory~\cite{mct} because 
the theory uniquely predicts the dynamical behaviour from static
density fluctuations. Gleim {\it et al.} argue that their finding 
of a quantitative agreement between SD and ND in a Lennard-Jones 
mixture is a nice confirmation
of this non-trivial mode-coupling prediction~\cite{gleim}. 
Szamel and Flenner~\cite{szamel2} 
confirmed this claim using BD, and argued further that 
even deviations from mode-coupling predictions are identical, 
a statement that was extended to below the mode-coupling
temperature by Berthier and Kob~\cite{ljmc}. 
In the present work we extend these findings to the case of silica
over a large range of temperatures, which goes far beyond 
the temperature regime where MCT can be applied. 
Therefore, we conclude that such an independence of the 
glassy dynamics of supercooled liquids to their 
microscopic dynamics, although
predicted by MCT, certainly has a much wider domain of validity 
than the theory itself. Finally, we note
that the deviations from MCT predictions observed in Fig.~\ref{comp}
cannot be attributed to coupling to currents
which are expressed in terms of particle velocities.
In our MC simulations we have no velocities,
so that avoiding the mode-coupling singularity is not due to 
the hydrodynamic effects pointed out in Ref.~\cite{previous} 
(see Ref.~\cite{more} for more recent theoretical viewpoints).

\begin{figure}
\psfig{file=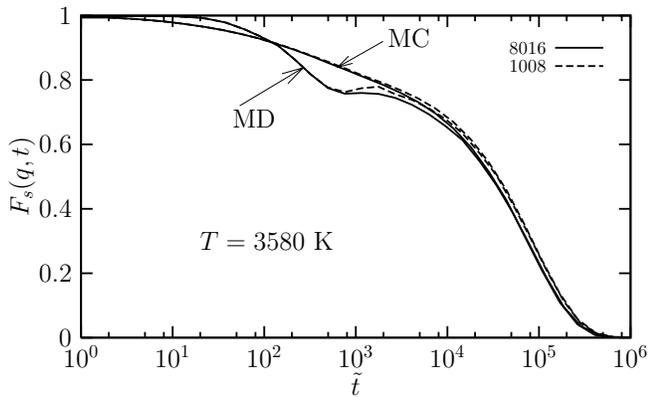,width=8.5cm}
\caption{\label{fss} Self-intermediate scattering function
for fixed $T$ and $q=1.7$~\AA$^{-1}$, obtained in MC and MD simulations
for two system sizes. The time axis in MD data is rescaled
by $t_0 = 0.31$ fs 
to obtain maximum overlap with MC results, and 
the same factor is  used for the two sizes. 
Larger systems relax faster and 
the amplitude of this finite size effect is the 
same for both dynamics.}
\end{figure}

The last comparison to ND we want to discuss  concerns the study 
of finite size effects. It was shown that the long-time dynamics
of silica is fairly sensitive to system size, and there are 
detectable differences when the number of particles is changed from 
1000 to 8000~\cite{hka,fss}. 
Such a large effect is not observed in more fragile 
materials~\cite{KA}. 
It was suggested that short-time thermal vibrations, stronger
in silica than in simpler models, are responsible for this system 
size dependence~\cite{hka,LW}. 
Therefore, it could be expected that by efficiently
suppressing these vibrations finite size effects should be reduced. 
But this is not what happens. In Fig.~\ref{fss}, we show 
self-intermediate scattering functions measured at $T=3580$~K
and $|{\bf q}|=1.7$~\AA$^{-1}$ in both ND and MC for two system sizes, 
$N=1008$ and $N=8016$ particles. Such data have been presented for
ND before~\cite{hka}, and our results agree with these earlier data. 
The amplitude of the vibrations observed for $t/t_0 = {\tilde t} 
\approx 10^3$ 
is smaller and the long time dynamics is faster when 
$N$ is larger. For MC we find that high-frequency vibrations
and the corresponding finite size effects are indeed suppressed, 
but the finite size effect for long-time relaxation, somewhat surprisingly, 
survives in our MC simulations, and can therefore not be attributed 
to high-frequency thermal vibrations. Recent studies
of the vibration spectrum and elastic properties at $T=0$ 
of amorphous media have suggested the existence of large-scale
structures~\cite{JL}: these objects are potential 
candidates to account for the size effect found at long times. 
It should then be  
explained how these spatial structures affect the long time dynamics, 
and why a finite size simulation box 
at the same time affects the absolute
value of the alpha-relaxation timescale but leaves unchanged
many of its detailed properties~\cite{fss2,heuer2}.     

\subsection{Mode-coupling analysis of dynamic correlators}

We now turn to a more detailed analysis of the shape and 
wave vector dependences of two-time correlation functions, 
revisiting in particular 
the mode-coupling analysis performed by Horbach and Kob
in Refs.~\cite{hk}. They argue that MCT can generally be applied
to describe their silica data, and attribute most of 
the deviations that they observe to short-time thermal vibrations
supposedly obscuring the ``true'' MCT behaviour. 
We are therefore in a position to verify if their hypothesis 
is correct. 

When applied to supercooled liquids, MCT formulates a series of detailed 
quantitative predictions regarding the time, wave vector, and temperature
dependences of two-time dynamical correlators close to the mode-coupling
singularity. In particular, MCT predicts that correlation functions
should indeed decay in the two-step manner reported in Fig.~\ref{fs}.
Moreover, for intermediate times corresponding to the plateau observed 
in correlation functions, an approximate 
equation can be derived  which
describes the correlator close enough to the plateau~\cite{mct}. 
The following
behaviour is then predicted, 
\be
F_s(q,t) \approx f_q + h_q F(t),
\label{betacorr}
\ee   
where $F(t)$ is the so-called $\beta$-correlator which 
is independent of the wave vector, and whose shape 
depends on a few parameters: the reduced distance from the 
mode-coupling temperature, $\epsilon = |T-T_c|/T_c$, and a parameter
describing the MCT critical exponents, $\lambda$. Once $\lambda$
is known various exponents $(a,b,\gamma)$ are known, which describe, in 
particular, the short-time behaviour of $F(t)$ when 
$F_s(q,t)$ approaches the plateau, $F(t) \sim t^{-a}$, and 
its long-time behaviour when leaving the plateau, 
$F(t) \sim t^b$. The exponent $\gamma$
was introduced in Eq.~(\ref{taumct}) 
and describes the temperature evolution
of the relaxation time $\tau_\alpha$.

\begin{figure}
\psfig{file=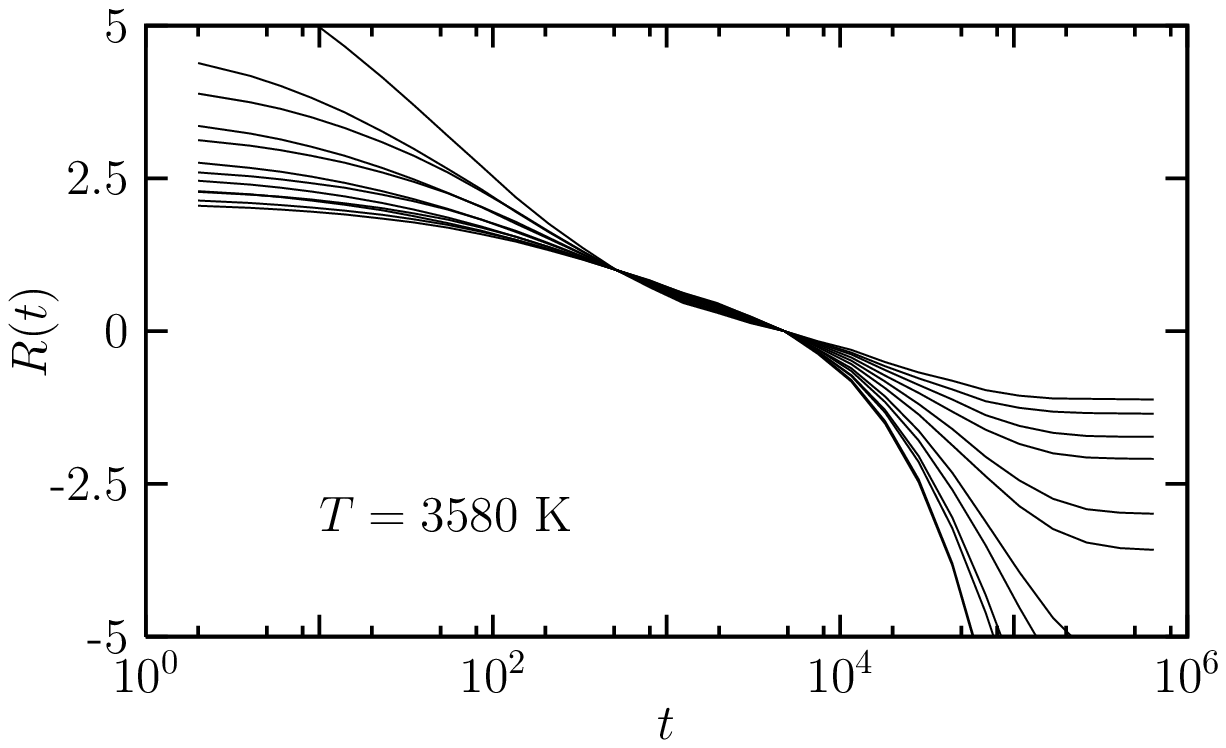,width=8.5cm}
\psfig{file=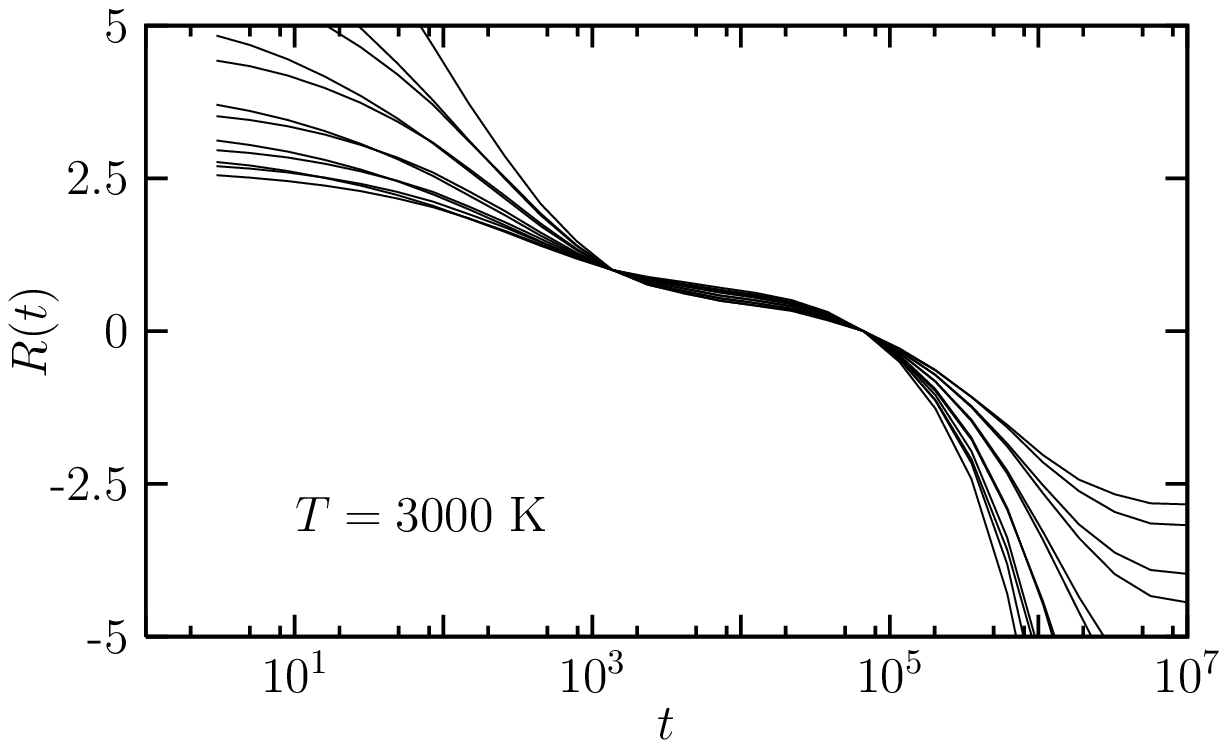,width=8.5cm}
\caption{\label{beta} 
Test of the factorization property, Eq.~(\ref{facteq})
using $F_s(q,t)$ from Si and O dynamics, and wave vectors between 
0.8 and 4 \AA$^{-1}$, for $T=3580$~K and 3000~K.
The data do not show collapse for times 
$t'<t<t''$, and factorization does not work very well.}
\end{figure}

Several properties follow from Eq.~(\ref{betacorr}). 
If one works at fixed temperature and varies the wave vector, 
the following quantity, 
\be
R(t) \equiv \frac{\phi(t)-\phi(t')}{\phi(t'')-\phi(t')} 
\approx \frac{F(t)-F(t')}{F(t'')-F(t')},
\label{facteq}
\ee
where $\phi(t)$ stands for a two-time correlation function, should 
become independent of $q$. 
In Eq.~(\ref{facteq}), $t'$ and $t''$ are two arbitrary times taken
in the plateau regime. This is called the ``factorization property''
in the language of MCT. We follow Ref.~\cite{hk} and show in Fig.~\ref{beta} 
the function $R(t)$ in 
Eq.~(\ref{facteq}) using self-intermediate scattering functions
for different $q$ and for different species (Si and O)
at fixed temperatures, $T=3580$~K and $T=3000$~K, choosing
times comparable to those reported in Ref.~\cite{hk}, namely 
$t''=82$ and $t'=760$ for $T=3580$~K, and 
$t''=1360$ and $t'=66700$ for $T=3000$~K.
Although the factorization property seemed to hold 
quite well in the ND data, 
this is no more the case for our MC data, and $R(t)$ retains 
a clear $q$ dependence between $t'$ and $t''$: no collapse of $R(t)$ 
can be seen in the regime $t'<t<t''$ in Fig.~\ref{beta}. 
The reason is clear from Fig.~\ref{fs}: due to 
thermal vibrations, the intermediate plateau was very flat in ND, 
but it has much
more structure in our MC data. It was therefore easier to collapse
the ND data in this regime than the present MC data for which 
a better agreement might have been expected.
In the case of the factorization property, the presence
of thermal vibrations 
in fact favours a positive reading of the data, which become much
less convincing when these vibrations are suppressed.  
Gleim and Kob had reached an opposite conclusion in the case of 
a Lennard-Jones system~\cite{gleim-kob}. They found 
that suppressing vibrations made the 
mode-coupling analysis of the beta-relaxation more 
convincing, suggesting that MCT describes 
the Lennard-Jones system more accurately than silica. 

\begin{figure}
\psfig{file=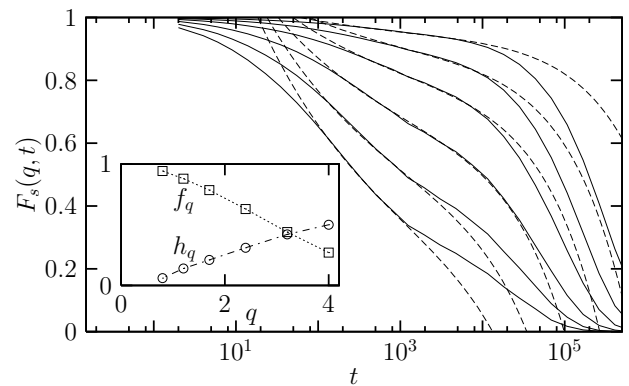,width=8.1cm}
\caption{\label{beta2} Self-intermediate scattering function
at fixed $T=3580$~K and various wave vectors, 
$q=0.8$, 1.2, 1.7, 2.4, 3.2, and 4~\AA$^{-1}$ (from right to left). 
Dashed lines show fits at intermediate times using Eq.~(\ref{betacorr}).
The inset shows the $q$-dependence of the fitting parameters
$h_q$ and $f_q$. Note that the time domains over which the fits 
apply shift with $q$.}
\end{figure}

Next we perform a test of the theory which had not been possible
with ND data. We investigate in detail if 
the behaviour predicted by Eq.~(\ref{betacorr}) is correct for both
short and long times. This test is not possible using ND because
the approach to the plateau is mainly ruled by thermal vibrations
(see for instance the ND data presented in Fig.~\ref{fss}).
In Fig.~\ref{beta2} we show that a ``critical decay'' does indeed 
show up when thermal vibrations are overdamped and no oscillations
can be seen. To check in more detail if this behaviour is indeed
in quantitative agreement with the MCT predictions, 
we fit the $F_s(q,t)$ data at $T=3580$~K, i.e. slightly 
above $T_c=3330$~K,
for several wave vectors $q$ using the
$\beta$-correlator obtained from numerical integration
of the mode-coupling equation.
To get the fits shown in Fig.~\ref{beta2} we have to fix
the distance to the mode-coupling temperature $\epsilon$ and 
the value $\lambda = 0.71$ both taken from Ref.~\cite{hk}, and yielding 
$a=0.32$, $b=0.62$ and $\gamma=2.35$.
Additionally we have to adjust the microscopic timescale.
Moreover, for each wave vector we have to fix $h_q$ and $f_q$ which 
respectively correspond to the amplitude of the $\beta$-correlator
and the height of the plateau in $F_s(q,t)$. 
Finally, there are two additional ``hidden'' free parameters 
in each of these fits: the somewhat arbitrarily chosen boundaries of 
the time domain where the fitting function describes the data.
We then get the fits shown with dashed lines in Fig.~\ref{beta2}, 
which are of a quality comparable to the ones 
usually found in the MCT literature~\cite{mct}. 
The parameters $h_q$ and $f_q$ are also 
shown in the inset of Fig.~\ref{beta2}, and behave 
qualitatively as in similar studies. Inspection of 
Fig.~\ref{beta2} reveals that the use of such freedom to fit the data
allows a qualitatively correct description of the data, although
clearly the time domain over which each wave vector is fit 
systematically shifts when $q$ changes, and we could 
not simultaneously fit the data at both small and large $q$ by fixing 
the time interval of the fit. This failure is consistent with the 
above finding that the factorization property is not satisfied. 

Therefore we conclude that 
MCT provides a qualitatively correct description of our data 
in the plateau regime, with no satisfying quantitative 
agreement, even in the absence of short-time thermal vibrations. 
One has therefore to argue that the data are taken too far from 
the transition for MCT to quantitatively apply to silica. However, since it 
is not possible to get data closer to the transition (recall that the 
transition does not exist), the domain of validity of the theory 
then would become vanishingly small.

\begin{figure}
\psfig{file=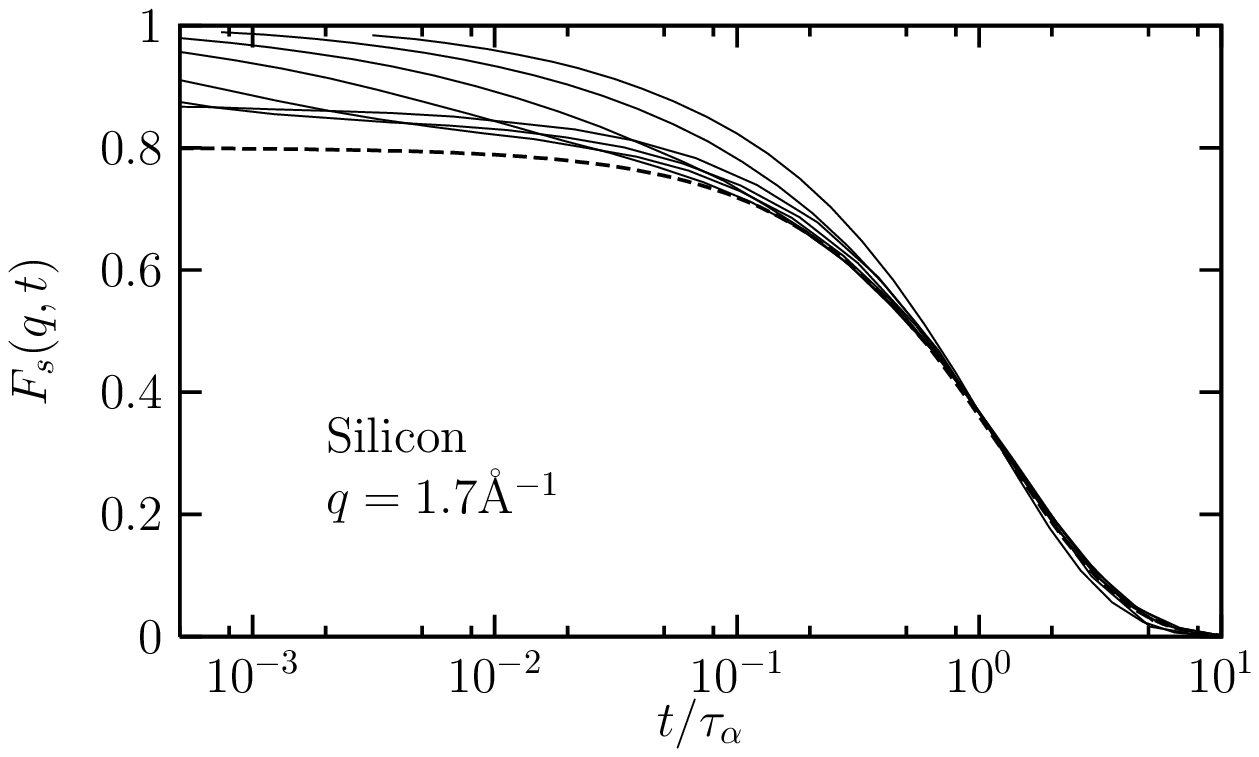,width=8.5cm}
\psfig{file=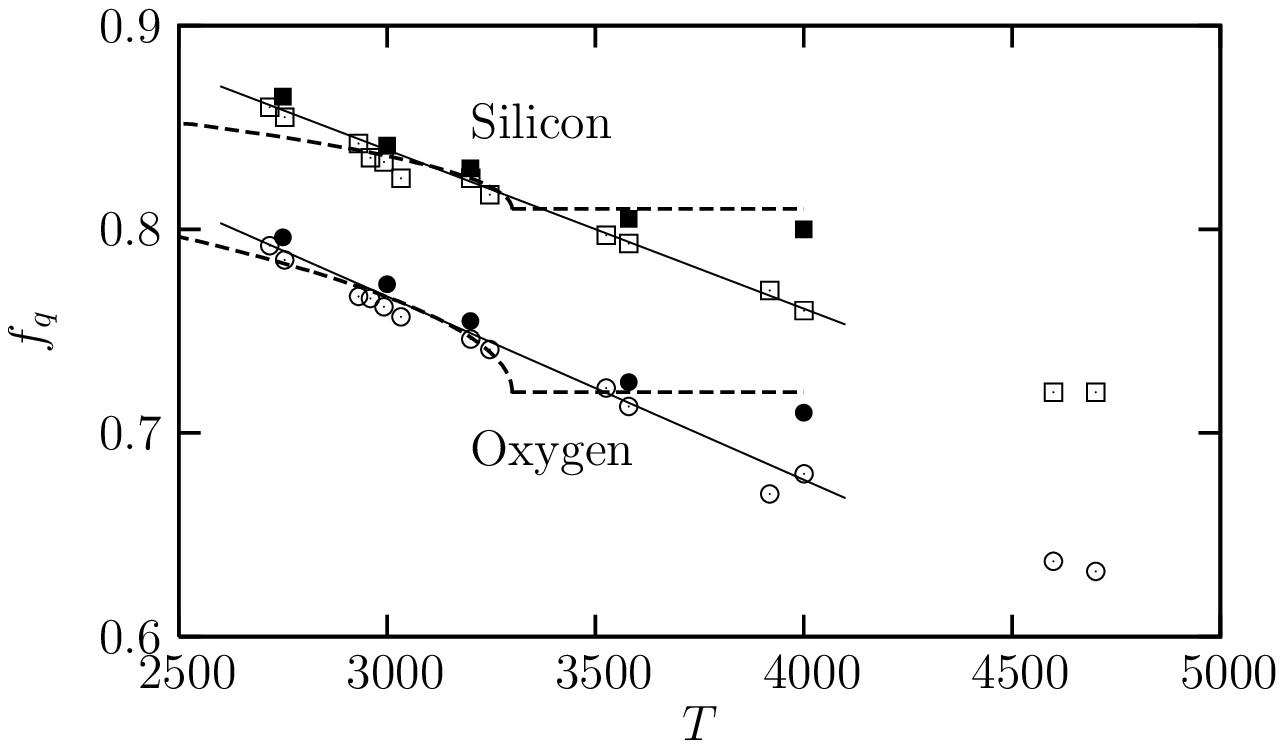,width=8.5cm}
\caption{\label{TTS} Top: Test of time-temperature superposition, 
Eq.~(\ref{tts}).
The dashed line is a stretched exponential function 
with $\beta=0.87$. Superposition holds at large rescaled times, but 
fails in the $\beta$-regime because the plateau height 
increases when $T$ decreases. Bottom: extracted plateau height
as a function of temperature fitted with a linear dependence (full line)
and with a square root singularity, Eq.~(\ref{sqrt}) (dashed line).
Open symbols are for ND, closed symbols for MC. No singular behaviour 
of $f_q$ is visible in either set of data.}
\end{figure}

We now turn to longer timescales and show 
in Fig.~\ref{TTS} a test of the time-temperature superposition 
prediction of the theory which states that correlators
at fixed $q$ but different
temperatures should scale as~\cite{mct} 
\be
F_s(q,t) \approx {\cal F}_q \left( \frac{t}{\tau_\alpha(q)} \right),
\label{tts}
\ee
where ${\cal F}_q(x) \approx f_q \exp (- x^{\beta(q)})$ and for times in 
the $\alpha$-regime. 
When high temperatures outside the glassy regime 
are discarded Eq.~(\ref{tts}) works correctly when
the scaling variable $t/\tau_\alpha$ is not too small, but
fails more strongly in the late $\beta$-regime. Scaling in the 
$\beta$-regime is often 
one of the most successful prediction of MCT, see e.g. Ref.~\cite{KA}.
In the present case, it could be argued to fail
because we are collapsing data at temperatures which 
are both above and below $T_c$. Indeed, below $T_c$ 
scaling in the $\beta$-regime is not expected anymore
because the height of the plateau, $f_q$ in Eq.~(\ref{betacorr}),
now becomes a temperature dependent quantity, 
with the following predicted singular behaviour~\cite{mct}:
\be
f_q(T) = f_q(T_c) + \alpha \sqrt{T_c - T}, \quad T \leq T_c,
\label{sqrt}
\ee
while $f_q(T \geq T_c) = f_q(T_c)$. 
The non-analytic behaviour of $f_q$ at $T_c$ is a further
characteristic feature of the mode-coupling singularity. 
Since we can easily take data for $T<T_c$ which are arguably not 
influenced by thermal vibrations, we can directly check for the presence 
of the square-root singularity, Eq.~(\ref{sqrt}). This is done in 
the bottom panel of Fig.~\ref{TTS}, 
where we also show data obtained from ND simulations.
That the latter are strongly influenced by thermal vibrations is clear, 
since they systematically lie below the MC data and have a stronger 
temperature dependence close to $T_c$. 
However, even the MC data clearly indicate that $f_q(T)$ is better described 
by a non-singular function of temperature, compatible 
with the simple linear behaviour expected to hold at 
very low temperatures.
The temperature dependence of the plateau height therefore explains
why time temperature superposition does not hold in the
late $\beta$-regime, but the linear temperature behaviour
indicates that there is no clear sign, from our data, 
of the existence of a ``true'' underlying singularity at $T_c$.

\section{Dynamic heterogeneity}
\label{dh}

Having established the ability of MC simulations
to efficiently reproduce the averaged 
slow dynamical behaviour observed in ND simulations, we now turn
to the study of the dynamic fluctuations around the average dynamical
behaviour, i.e. to dynamic heterogeneity. 

Dynamic fluctuations can be studied through a  
four-point susceptibility, $\chi_4(t)$, which 
quantifies the strength of the spontaneous fluctuations 
around the average dynamics by their variance, 
\begin{equation} 
\chi_4(t) = N_\alpha \left[  \langle f_s^2({\bf q}, t) 
\rangle - F_s^2({ q}, t) \right], 
\label{chi4lj} 
\end{equation} 
where 
\be
f_s({\bf q},t) = \frac{1}{N_\alpha} \sum_{i=1}^{N_\alpha}  
\cos ({\bf q} 
\cdot [{\bf r}_i(t) - {\bf r}_i(0)] ),
\ee 
represents
the real part of the instantaneous value of 
the self-intermediate scattering function, 
so that $F_s({ q},t) = \langle f_s({\bf q},t) \rangle$.
As shown by Eq.~(\ref{chi4lj}),
$\chi_4(t)$ will be large if run-to-run fluctuations
of the self-intermediate scattering functions 
averaged in large but finite volume,
are large. This is
the case when the local dynamics becomes
spatially correlated, as already discussed in several 
papers~\cite{FP,silvio2,glotzer,lacevic,toni,mayer,berthier}.

\begin{figure}
\psfig{file=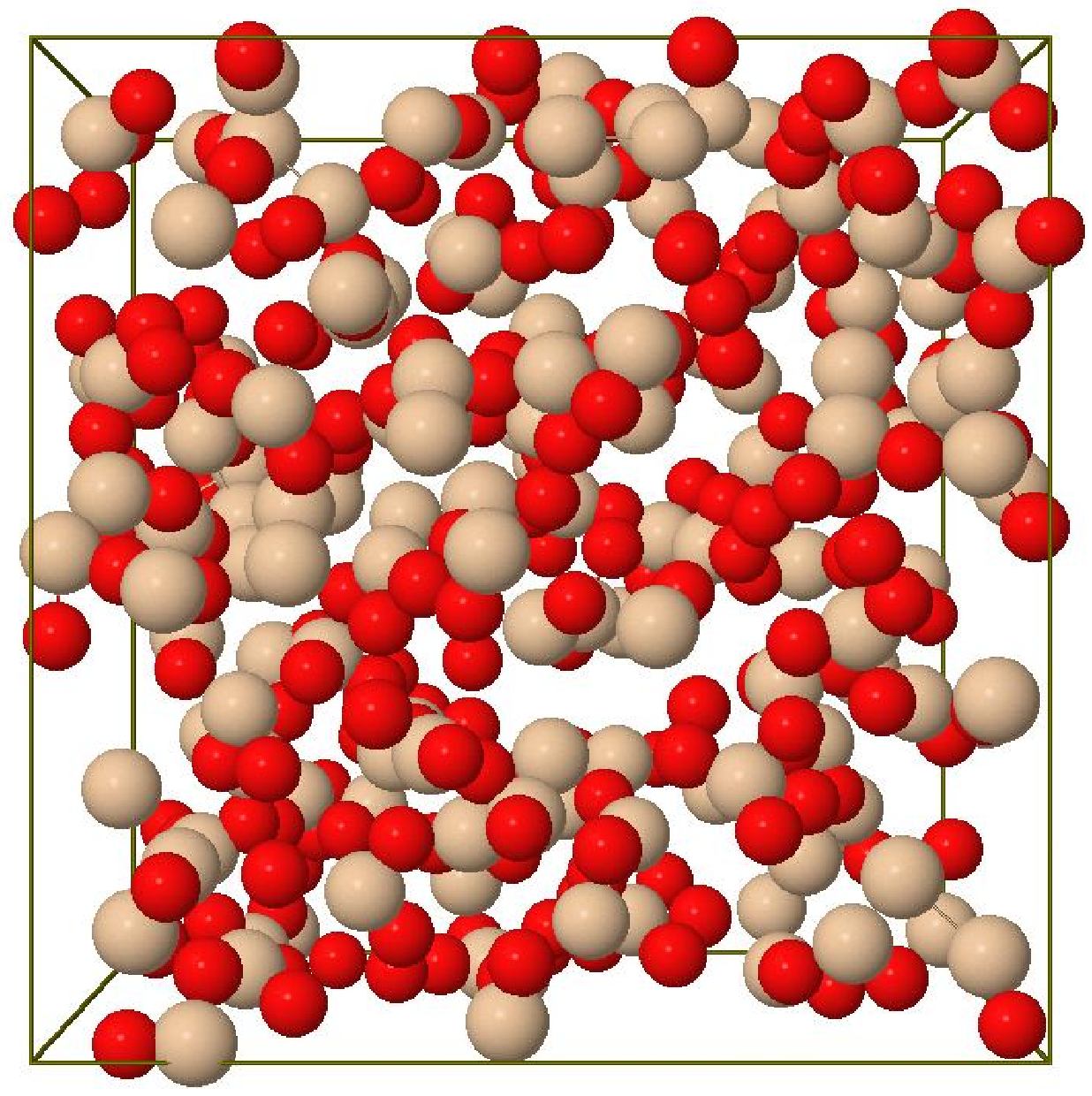,width=5.3cm}
\psfig{file=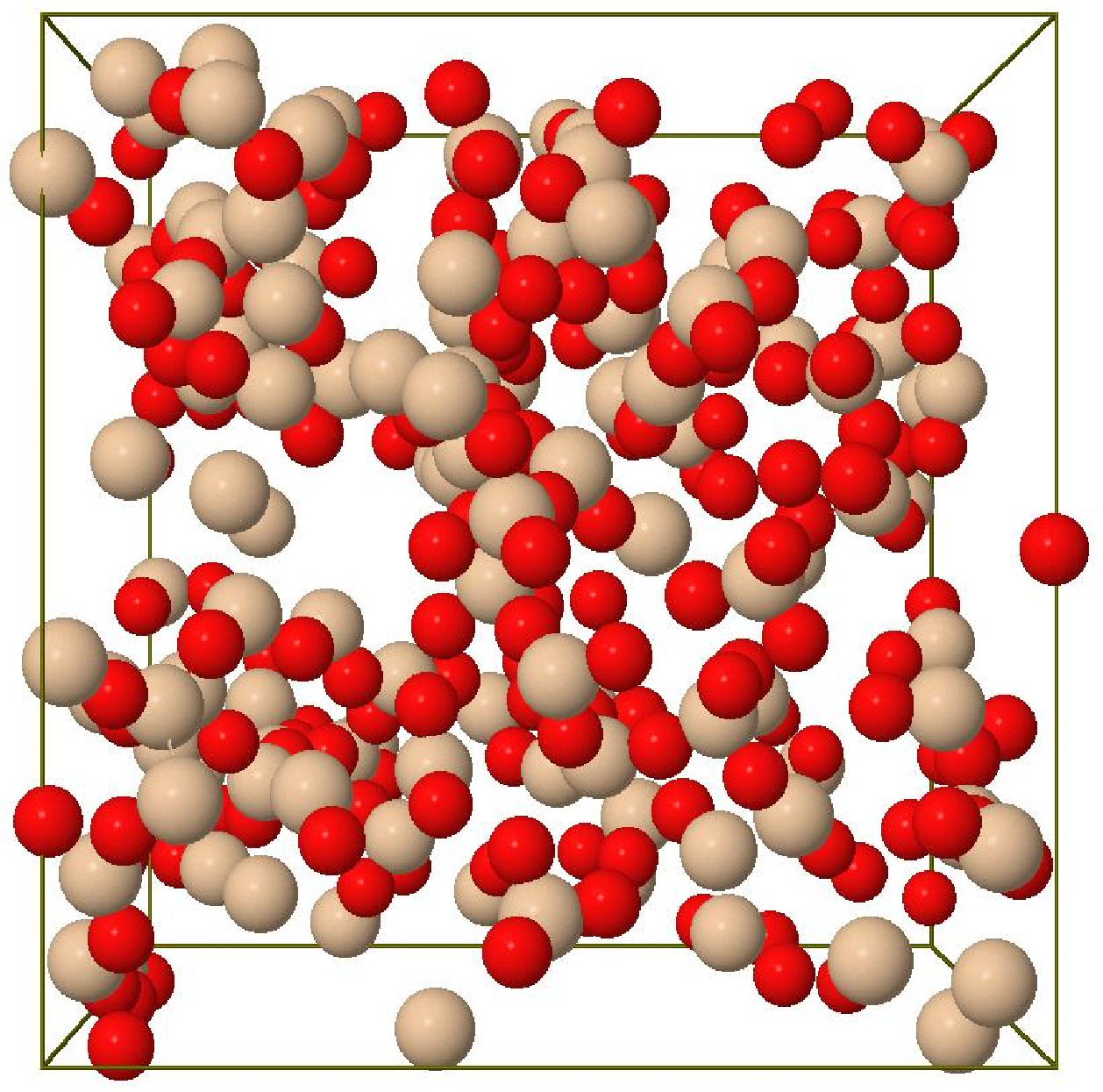,width=5.3cm}
\psfig{file=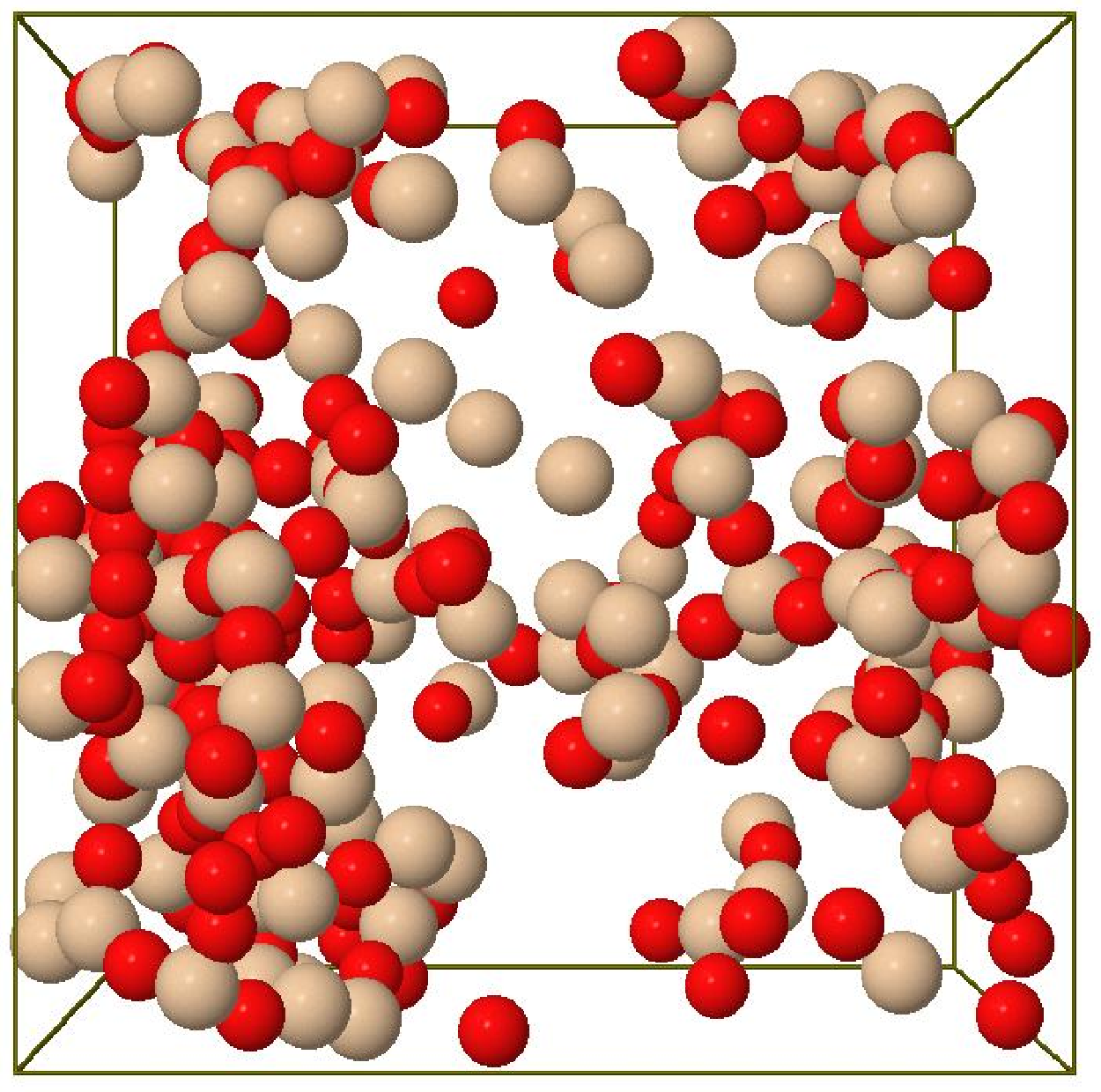,width=5.3cm}
\caption{\label{snap} Snapshots of dynamic heterogeneity at $T=6100$,
3580 and  3000~K (from top to bottom). The snapshot presents 
particles which, in a particular run at a particular temperature 
have been slower than the average, and have therefore 
large, positive values of $\delta f_i({\bf q},t \approx \tau_\alpha)$ 
defined in Eq.~(\ref{snapeq}). Light colour is used for Si, dark for O.
Slow particles tend to cluster in space 
on increasingly larger lengthscales when $T$ decreases.}
\end{figure}

What $\chi_4(t)$ captures is information on the spatial
structure of the spontaneous fluctuations of the dynamics around their 
average. We define 
$f_i({\bf q},t) = \cos ({\bf q} 
\cdot [{\bf r}_i(t) - {\bf r}_i(0)] )$, the contribution 
of particle $i$ to the instantaneous value of $F_s(q,t)$, 
and 
\be 
\delta f_i({\bf q},t) = f_i({\bf q},t) - F_s(q,t),
\label{snapeq}
\ee 
its fluctuating 
part. Then $\chi_4(t)$ can be rewritten in the suggestive form, 
\be
\chi_4(t) = \rho \int d{\bf r} \left\langle 
\sum_{i,j} \delta f_i({\bf q},t)  
\delta f_j({\bf q},t) \delta ( {\bf r} - [{\bf r}_i(0) - {\bf r}_j(0)] )
\right\rangle ,
\label{volume}
\ee
where subtleties related to the exchange between thermodynamic limit
and thermal average are discussed below. 
Therefore $\chi_4(t)$ is the volume integral of the spatial
correlator between local fluctuations of the dynamical behaviour 
of the particles. It gets larger when the spatial range of these 
correlations increases. 

To get a feeling of how 
these fluctuations look like in real space,
we present snapshots at different temperatures 
in Fig.~\ref{snap}. To build these snapshots
we show, for a given run at a given temperature, those particles 
for which the fluctuating quantity $\delta f_i({\bf q},t \approx 
\tau_\alpha)$, is 
positive and larger than a given threshold, which we choose
close to 1/2 for graphical convenience (this leads to about 
1/3 of the particles being shown, and clearer snapshots).
The shown particles are therefore slower than the average
for this particular run. The evolution of the snapshots between 6100~K and 
3000~K clearly reveals the tendency for slow particles to cluster in space,
revealing the growth of the lengthscale of  kinetic heterogeneities. 
We should note, however, that the clusters shown here are not 
macroscopic objects even at the lowest temperature studied. 
Moreover, similar snapshots in Lennard-Jones 
systems reveal more clearly the tendency we seek to 
illustrate~\cite{berthier}. 
We interpret this as a further qualitative indication that dynamic
heterogeneity is less pronounced in this Arrheniusly relaxing 
material than in more fragile Lennard-Jones systems.

\begin{figure}
\begin{center}
\psfig{file=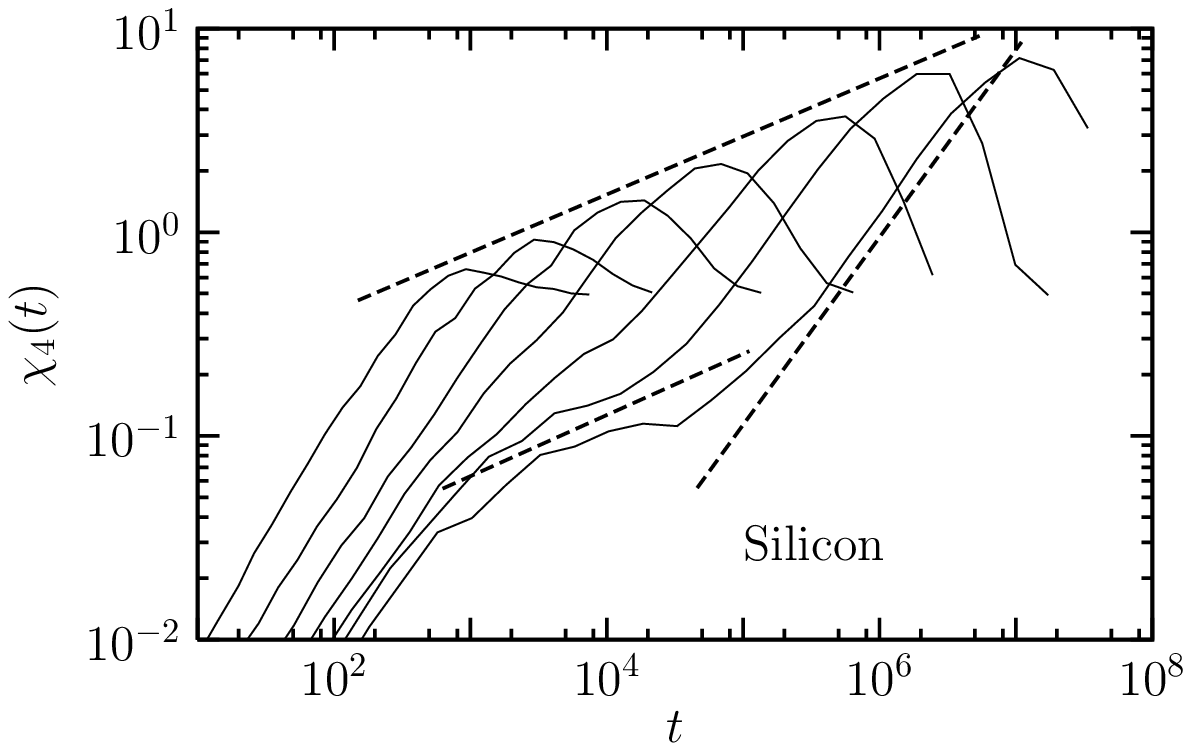,width=8.5cm}
\psfig{file=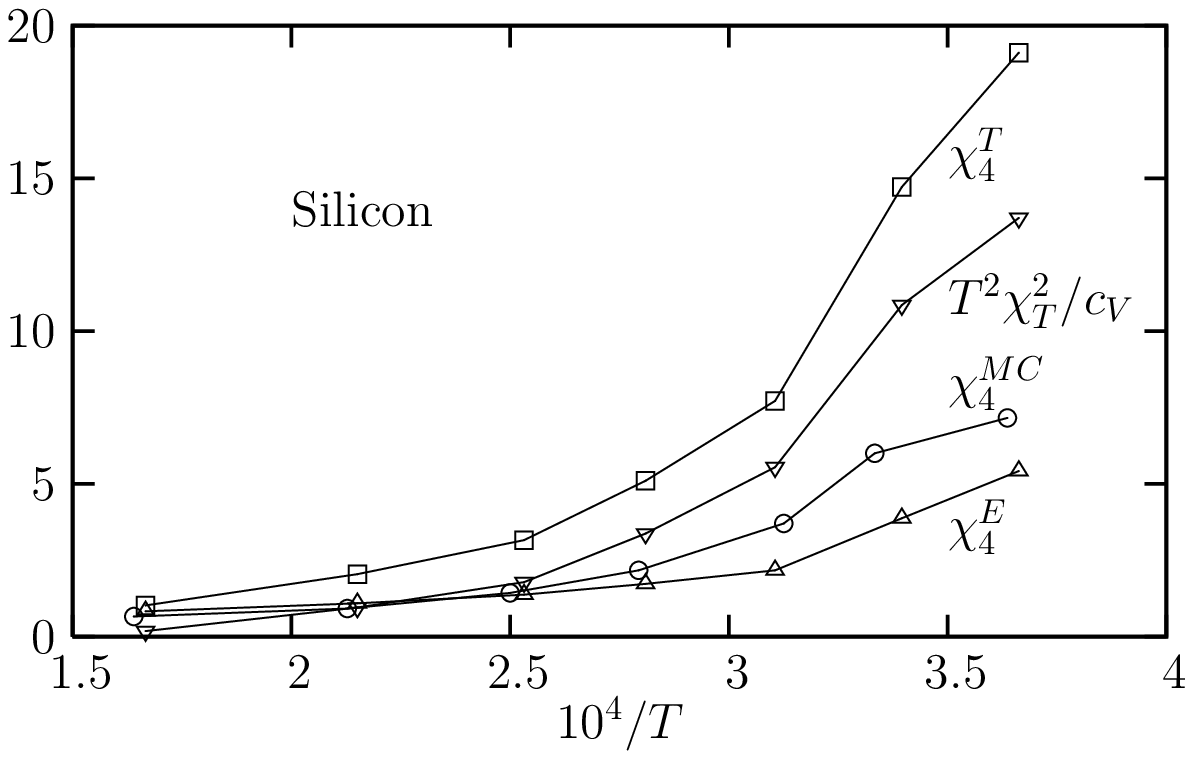,width=8.5cm}
\end{center}
\caption{\label{chi4} Top: Four-point susceptibility, Eq.~(\ref{chi4lj}),
for the same temperatures as in Fig.~\ref{fs}, decreasing from 
left to right. The low temperature data at $T=2750$~K 
are fitted with two power laws shown as 
dashed lines with exponents $0.3$ and $0.92$ at short and large times,
respectively. The envelope of the maxima is fit with an exponent 0.285.
Bottom: temperature evolution of the maxima in various 
dynamic susceptibilities.}
\end{figure}

We turn to more quantitative measures of dynamic heterogeneity and
show the time dependence of the dynamic susceptibility
$\chi_4(t)$ obtained from our MC simulations for various temperatures 
in Fig.~\ref{chi4}. Similar data are obtained for Si and O, 
and we only present the former.
As predicted theoretically 
in Ref.~\cite{toni} we find that $\chi_4(t)$ presents
a complex time evolution, closely related to 
the time evolution of the self-intermediate scattering function.
Overall, $\chi_4(t)$ is small at both small and large 
times when dynamic fluctuations are small. There is therefore 
a clear maximum observed for times comparable to 
$\tau_\alpha$, where fluctuations are most pronounced. The position
of the maximum then shifts to larger times when 
temperature is decreased, tracking the alpha-relaxation timescale.
The most important physical information revealed by these curves is the
fact that the amplitude of the peak grows when the temperature 
decreases. This is direct evidence, recall Eq.~(\ref{volume}), 
that spatial correlations grow 
when the glass transition is approached.

The two-step decay of the self-intermediate scattering function
translates into a two-power law regime for $\chi_4(t)$ 
approaching its maximum. We have fitted these power laws,
$\chi_4(t) \sim t^a$, followed by $\chi_4(t) \sim t^b$ with the 
exponents $a =0.3$ and $b=0.92$ in Fig.~\ref{chi4}. We 
have intentionally used the notation $a$ and $b$ for these 
exponents which are predicted, within mode-coupling theory, 
to be equal to the standard exponents also describing
the time dependence of intermediate scattering functions~\cite{toni,II}. 
Our findings are in reasonable agreement with
values for $a$ and $b$ discussed above, although the $b$-value
is about 50\% too large. Moreover, a two-power law regime
is only observed for $T<T_c$, where MCT does not apply anymore. 
We note that the $b$-value is predicted to be $b=2$ 
from the perspective of modelling strong glass-formers using 
kinetically constrained models with Arrhenius behaviour~\cite{steve2}; 
this prediction is clearly incorrect for BKS silica~\cite{toni,II}.

We then focus on the amplitude of the dynamic susceptibility 
at its maximum and follow its temperature evolution in Fig.~\ref{chi4}.
As suggested by the snapshots shown in Fig.~\ref{snap}, we confirm
that spatial correlations increase when $T$ decreases, as $\chi_4$ gets 
larger at low temperatures. The temperature evolution 
of the peak was discussed in Ref.~\cite{II}. Both MCT
and kinetically constrained models 
strongly overestimate the temperature evolution 
of $\chi_4$ at its peak value, as emphasized already in Ref.~\cite{II}.
Finally, we note that the typical values 
observed for the peak of $\chi_4$ at low temperatures are 
significantly smaller than those observed for more fragile 
Lennard-Jones systems,  suggesting once more that dynamic
heterogeneity is less pronounced  in strong glass-forming
materials. 

This comparison is also useful to discuss the possibility 
of finite size effects on the present $\chi_4$ data. If computed 
in a simulation box which is too small, the dynamic 
susceptibility takes values that are too small~\cite{fssprl}. 
Our data indicate that no saturation of the maximum value
of $\chi_4(t)$ is reached, and the values we find 
of smaller than the ones found in a Lennard-Jones system with 
a comparable system size and for which a detailed 
search for possible finite size effects was performed~\cite{I}.
We believe therefore that our results are not affected 
by finite size effects.

We then compare these results to the ones obtained using 
Newtonian dynamics in the same system. 
In that case, care must be taken of the order  
at which the thermodynamic limit and the thermal average are taken in 
Eq.~(\ref{volume}). Indeed when ND is used, the dynamics strictly conserves
the energy during the simulation and thermal averages are then performed
in the microcanonical ensemble, and $\chi_4^E$ is measured. 
To measure $\chi_4^T$ in the canonical ensemble for ND, an additional 
contribution must be added, which takes into account 
the amount of spontaneous fluctuations which are due to 
energy fluctuations~\cite{science}, 
\begin{equation}
\chi_4^{T}(t) - \chi_4^{E}(t) = \frac{T^2}{c_V} \left(
\frac{\partial F_s({ q},t)}{\partial T} \right)^2 \equiv \frac{T^2}{c_V}
\chi_T(t),
\label{chiT}
\end{equation}
where $c_V$ is the constant volume specific heat expressed in 
$k_B$ units.
The results for $\chi_4^T$ and $\chi_4^E$ obtained from ND, 
and the difference term in Eq.~(\ref{chiT}), 
are all presented in Fig.~\ref{chi4}. 
We find that the MC results for $\chi_4$ lie closer to the 
microcanonical results obtained from ND, while the 
canonical fluctuations are significantly larger, due to the 
large contribution of the right hand side in Eq.~(\ref{chiT}). 
This is at first sight contrary to the intuition
that MC simulations are thermostatted and should be 
a fair representation of canonical averages in ND. 
But this is not what happens. As discussed
in Refs.~\cite{I,II}, a major role is played by
conservation laws for energy and density when 
dynamic fluctuations are measured. 
In the case of energy conservation the mechanism can
be physically understood as follows. For a rearrangement 
to take place in the liquid,
the system has to locally cross an energy barrier. If dynamics conserves the
energy, particles involved in the rearrangement must borrow 
energy to the neighboring particles. This `cooperativity' 
might be unnecessary if  energy can be locally supplied
to the particles by an external heat bath, as in MC simulations. 
Conservation laws, therefore, might
induce dynamic correlations between particles and dynamic fluctuations 
can be different when changing from Newtonian, energy conserving 
dynamics to a stochastic, thermostatted dynamics. 
With hindsight, this is 
not such a surprising result. The specific heat, after all,
also behaves differently in different statistical ensembles.
The ensemble dependence and dependence upon the microscopic dynamics
of dynamic susceptibilities in supercooled liquids
are the main subjects of two recent papers~\cite{I,II}.
Our results for silica quantitatively agree
with the theoretical analysis they contain, and with 
the corresponding numerical 
results obtained in Lennard-Jones systems.

There is an experimentally extremely relevant consequence of 
these findings~\cite{science,cecile}. 
As shown in 
Fig.~\ref{chi4}, the difference
between the microcanonical and canonical values of the dynamic 
fluctuations in ND represents in fact the major
contribution to $\chi_4^{T}$, meaning
that the term $\chi_4^{E}$ can be safely neglected in Eq.~(\ref{chiT}).
Since the right hand side of (\ref{chiT}) is 
more easily accessible in an experiment than $\chi_4^T$ itself, 
Eq.~(\ref{chiT}) opens the possibility of an experimental
estimate of the four-point susceptibility. This finding, 
and its experimental application to supercooled glycerol and 
hard sphere colloids, 
constitute the central result of Ref.~\cite{science}, while more data
are presented in Ref.~\cite{cecile}.

\section{Conclusion}
\label{conclusion}

We have implemented a standard Monte Carlo algorithm to study the 
slow dynamics of the well-known BKS model for silica in the temperature
range from 6100~K to 2750~K. Our results clearly establish 
that Monte Carlo simulations can be used to study the dynamics 
of silica because quantitative agreement is found with results
from Newtonian dynamics for the same potential, apart at very short times
where thermal vibrations are efficiently suppressed by the Monte Carlo
algorithm. The agreement between the two dynamics 
is by no means trivial and constitutes an important result of the present 
study. This suggests that Monte Carlo simulations
constitute a useful and efficient tool to study also the nonequilibrium aging 
dynamics of glass-forming liquids, a line of research initiated in 
Ref.~\cite{berthier2}.

Since dynamical correlations are not affected by short-time vibrations, 
we have been able to revisit the mode-coupling analysis 
initially performed in Ref.~\cite{hk}. We find that mode-coupling theory
accounts for the qualitative features of the data quite well, but 
the detailed, quantitative predictions made by the theory were shown 
to fail: correlation functions close to the plateau 
do not follow the behaviour predicted for the MCT $\beta$-correlator, 
time-temperature superposition only holds at very large times but fails 
at smaller times because the plateau in correlation function is strongly
temperature dependent, a dependence which does not follow the 
singular behaviour predicted by MCT. Moreover, the temperature 
regime where the theory can supposedly be applied is found to be 
at most 1 decade when only 
the temperature evolution of relaxation timescales
is considered. Furthermore, we have argued that the motivation 
to analyze silica data in terms of MCT, a ``fragile to strong'' crossover, 
can in fact be more simply accounted for in terms 
of a crossover between two distinct Arrhenius regimes occurring close to 
$T \approx 4000$~K. Overall these results suggest
a negative answer to the question: is there
any convincing evidence of an avoided
mode-coupling singularity in silica? 

We have finally analyzed dynamic heterogeneity in silica. We find that
the dynamics is indeed spatially heterogeneous, and spatial 
correlation of the local dynamical behaviour was shown to increase 
when temperature decreases. We also found that all indicators 
of dynamically heterogeneous dynamics such as decoupling and four-point
dynamic susceptibilities, suggest that the effects are less pronounced
in silica than in more fragile glass-forming materials, but do not
seem qualitatively different. 

The most natural interpretation is that 
strong and fragile materials in fact belong to 
the same class of materials, where the effects of dynamic heterogeneity
could become less pronounced, but definitely non-zero, 
for materials with lower fragility. 
This suggests that it could be incorrect to
assume that strong materials belong to a different universality class
from fragile ones, as studies of kinetically constrained models
with different fragilities would suggest~\cite{steve2,KCM}, 
and they should rather seat at the end 
of the spectrum of fragile systems. It seems however similarly incorrect
to consider that strong materials are ``trivial'' because an Arrhenius 
behaviour can be explained from simple thermal activation over a fixed
energy barrier corresponding to a local, non-cooperative event. Our  results
show that this is not a correct representation of the physics of strong 
glass-formers either. Convincingly incorporating 
fragility into current theories of the glass transition while 
simultaneously giving it a microscopic interpretation
remains therefore an important challenge. 

\begin{acknowledgments}
This work emerged from collaboration with G. Biroli, J.-P. Bouchaud,
W. Kob, K. Miyazaki, and D. Reichman~\cite{I,II}.
D. Reichman suggested to revisit the MCT analysis of BKS silica,
W. Kob helped analyzing and interpreting the results, and A. Heuer 
made useful comments on the preprint.
This work has been supported in part by Joint Theory Institute at 
Argonne National Laboratory and the University of Chicago. 
\end{acknowledgments}

\end{document}